\documentclass[%
superscriptaddress,
twocolumn,
prb,
floatfix,
]{revtex4-2}

\usepackage{graphicx}
\usepackage{dcolumn}
\usepackage{bm}
\usepackage{ulem}
\usepackage{xcolor}
\usepackage{amsmath} 
\usepackage{hyperref}
\usepackage{multirow}

\definecolor{newtext}{RGB}{0, 100, 100}
\definecolor{AK}{RGB}{250, 150,0}

\begin{document}

\preprint{APS/123-QED}

\title{
Nonreciprocal collective magnetostatic wave modes in geometrically asymmetric bilayer structure with nonmagnetic spacer}

\author{P. I. Gerevenkov}
\email{petr.gerevenkov@mail.ioffe.ru}
\homepage{http://www.ioffe.ru/ferrolab/}
 \affiliation{Ioffe Institute, 194021 St. Petersburg, Russia}
\author{V. D. Bessonov}%
\affiliation{%
 M.N. Mikheev Institute of Metal Physics, UB of RAS, 620108 Ekaterinburg, Russia
}%
\author{V. S. Teplov}%
\affiliation{%
 M.N. Mikheev Institute of Metal Physics, UB of RAS, 620108 Ekaterinburg, Russia
}%
 \author{A. V. Telegin}%
\affiliation{%
 M.N. Mikheev Institute of Metal Physics, UB of RAS, 620108 Ekaterinburg, Russia
}%
 \author{A. M. Kalashnikova}
  \affiliation{Ioffe Institute, 194021 St. Petersburg, Russia}
\author{N. E. Khokhlov}
 \affiliation{Ioffe Institute, 194021 St. Petersburg, Russia}

\date{\today}

\begin{abstract} 
Nonreciprocity, i.e. inequivalence in amplitudes and frequencies of spin waves propagating in opposite directions, is a key property underlying functionality in prospective magnonic devices. 
Here we demonstrate experimentally and theoretically a simple approach to induce frequency nonreciprocity in a magnetostatically coupled ferromagnetic bilayer structure with a nonmagnetic spacer by its geometrical asymmetry.
Using Brillouin light scattering, we show the formation of two collective spin wave modes in Fe$_{81}$Ga$_{19}$/Cu/Fe$_{81}$Ga$_{19}$ structure with different thicknesses of ferromagnetic layers.
Experimental reconstruction and theoretical modeling of the dispersions of acoustic and optical collective spin wave modes reveal that both possess nonreciprocity reaching several percent at the wavenumber of $22~\cdot~10^4$\,rad/cm. 
The analysis demonstrates that the shift of the amplitudes of counter-propagating coupled modes towards either of the layers is responsible for the nonreciprocity because of the pronounced dependence of spin wave frequency on the layers thickness.
The proposed approach enables the design of multilayered ferromagnetic structures with a given spin wave dispersion for magnonic logic gates.
\end{abstract}

\maketitle


\section*{Introduction}
Nonreciprocity, i.e. inequivalence of a medium properties sensed by a (quasi)particle traversing it in positive and negative directions, underlies the operation of indispensable elements of modern electronic, microwave techniques, and optics.
Limitations of transistor density and power consumption associated with conventional electron-charge technologies promote other quesiparticles as information carriers for next generation of computing.
An emerging competitor of electronics, magnonics~\cite{mahmoud2020introduction, li2020hybrid} employs spin waves and their quasi-particles, magnons, to encode, carry, and process the information.
Thus, magnonics is eager for the development of basic nonreciprocal elements, such as spin wave diode~\cite{szulc2020spin, grassi2020slow}, circulator~\cite{szulc2020spin}, half-adder~\cite{wang2020magnonic}, etc.
The clear advantage of magnonics is that nonreciprocity of spin waves' amplitudes is intrinsic for magnetic thin films, as was recognized already in the 1960s~\cite{damon1961magnetostatic}.
However, the functionality and tunability of nonreciprocal magnonic elements~\cite{Barman_2021,nikitov2015magnonics} broaden significantly if they support frequency nonreciprocity as well, when waves of the same wavenumber traveling in opposite directions possess different frequencies~\cite{camley1987nonreciprocal}.
These require finding media or designing artificial structures possessing strong frequency nonreciprocity for spin waves while being compatible with miniaturization and maintaining low losses. 
Intrinsic frequency nonreciprocity of the spin waves requires that the medium possess chirality~\cite{seki2016magnetochiral}, which is lacking in most materials relevant for magnonics applications.
Therefore, symmetry breaking of various origins in thin films and heterostructures is now seen as the most promising way to realize non-reciprocity of spin waves~\cite{udvardi2009chiral, zakeri2010asymmetric,verba2013conditions, gallardo2019spin,Dieny:RMP2017, kruglyak2021chiral}.

It has recently been recognized that the collective behavior of spin waves in coupled waveguides, magnonic crystals, etc. offers extended possibility to design and tune dispersion properties.
This owns to the fact that even small change in properties of one of the element of such a structure may result in drastic changes in behavior of coupled spin wave modes~\cite{krawczyk2014review, carlotti1999brillouin}. 
In this respect, heterostructures with two coupled magnetic layers have recently attracted great interest due to their ability to support collective magnetization dynamics and to the possibility of controlling it by independently adjusting characteristics of individual constituents~\cite{grachev2021strain} and coupling between them~\cite{andersson2010exchange, heinrich1993ultrathin}.
In contrast to the widely considered nonreciprocity associated with interfacial anisotropy~\cite{gladii2016frequency} and / or the Dzyaloshinskii-Moriya interaction~\cite{gallardo2019flat, franco2020enhancement, moon2013spin}, the difference in the dispersion of collective magnetostatic spin wave (MSW) modes propagating in opposite directions in multilayered structures also relies on the interaction between layers.
In~\cite{shiota2020tunable, sud2020tunable, gallardo2019reconfigurable, ishibashi2020switchable}, nonreciprocity is demonstrated in synthetic antiferromagnets, systems with antiferromagnetic ordering of ferromagnetic layers due to indirect exchange interaction.
Nonreciprocity has been demonstrated in systems of exchange-coupled layers with a parallel orientation of magnetization~\cite{grassi2020slow,song2020backward,odintsov2022nonreciprocal}. 
On the other hand, in~\cite{song2020backward, kub2021nonreciprocal}, it was shown that multilayered systems characterized by differences in the magnetic parameters of their constituents, e.g. saturation magnetization and anisotropy, also support the nonreciprocity of MSW, which is based on asymmetry in magnetic parameters.
The antiparallel orientation of magnetizations in dipole-coupled layers also serves as a source of nonreciprocity~\cite{gallardo2019reconfigurable, franco2020enhancement}. 

In this article, we propose to realize nonreciprocal MSW propagation in a ferromagnetic bilayer structure, where the two dipole-coupled ferromagnetic layers possess similar magnetic properties, but have different thicknesses.
The structure exhibits a parallel orientation of the magnetizations of the layers.
Using Brillioun light scattering, we examine the dispersion of thermal MSWs in the bilayer structure and reveal the formation of collective acoustic (in-phase) and optical (out-of-phase)  modes.
Both modes demonstrate considerable nonreciprocity reaching several percent at a wavenumber of $22\cdot10^4$\,rad cm$^{-1}$. 
Using a theoretical model which takes into account dipolar, exchange, and anisotropy contributions to spin wave dispersion, we show that, along with the difference in layers' magnetic parameters, the leading contribution to the nonreciprocity comes from the difference in the spin wave frequencies of the layers due to their different thicknesses. 
The thickness effect is pronounced in films with a thickness of the order of 10~nm and can be additionally enhanced due to interfacial spin pinning.
We further suggest the conditions for maximizing the nonreciprocity, which gives an optimal geometrical asymmetry defined by the relation between the layer thicknesses and the pinning conditions at the interfaces.
Such artificially induced nonreciprocity gives the opportunity to design magnonic logic gates based on their geometrical parameters rather than on the properties of the materials.

\section*{Sample and methods}
\begin{figure*}
\centering
\includegraphics[width= 0.85 \linewidth]{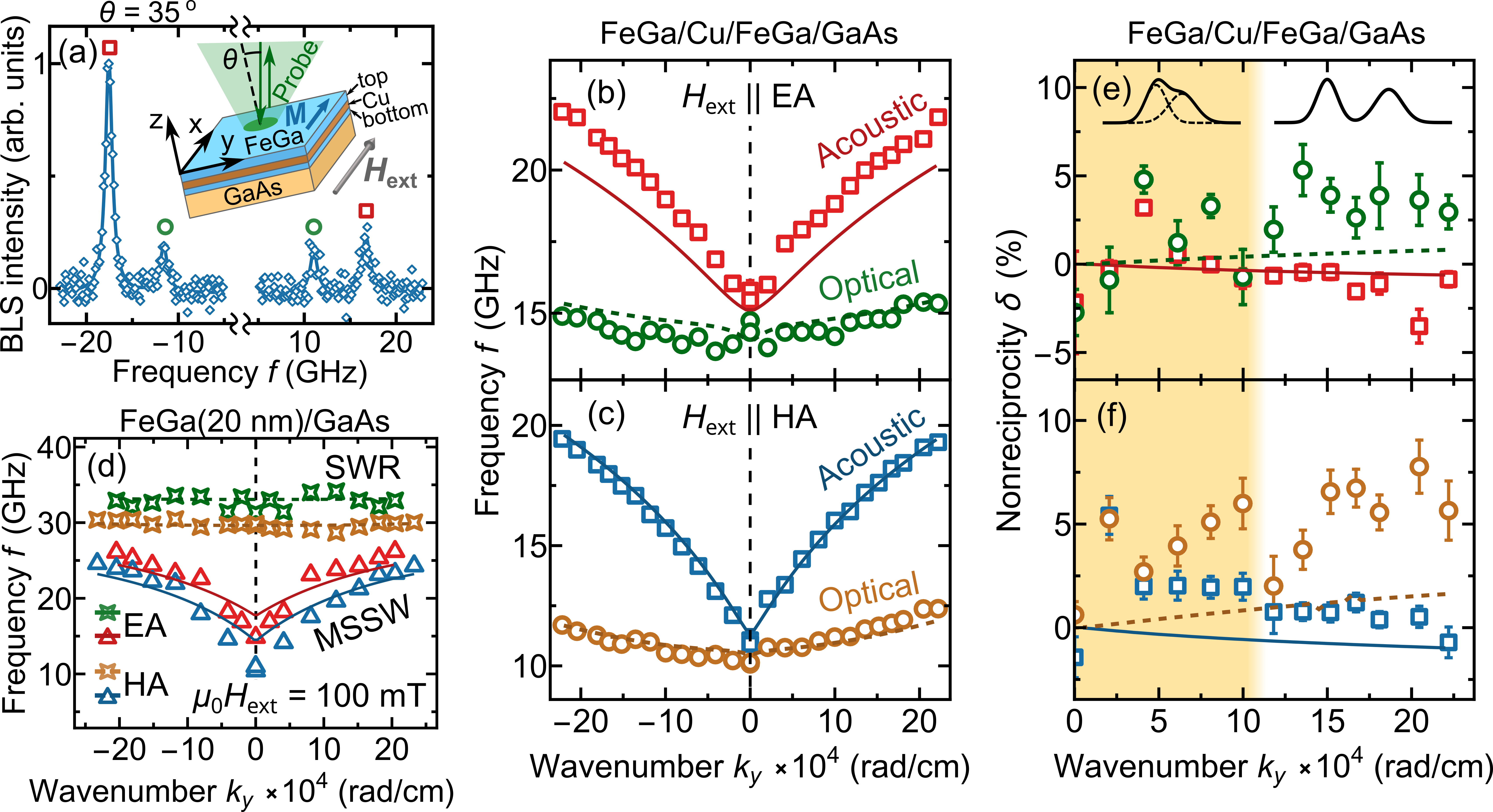}
\caption{\label{fig:3Lay}
(a) Experimental BLS signal (diamonds) for Stokes and anti-Stokes parts of the spectrum for bilayer sample at $k_y =  13.5\cdot10^4$\,rad~cm$^{-1}$ (tilted angle $\theta = 35$\,$^\circ$). 
Solid line shows best fit with Lorenz peaks.
Insert: geometry of the experiment.
(b and c) Thermal magnons dispersions in bilayer sample for $\mathrm{{\bf H}_{ext}}$ applied along easy and hard magnetization directions, respectively.
(d) Thermal magnons dispersions for magnetization along easy and hard axes in the single galfenol film.  
(e and f) Nonreciprocity parameter $\delta$ of MSWs vs. wavenumber $k_y$ determined using data from (b and c), respectively.
Lines in (b-f) show the data from the model (see Sec.~I Supp.Mat.~\cite{Supplemental}).
The color area on panel (e and f) separates the wavenumber ranges at which the resolution of the peaks in the spectrum is achieved or not (see sketch at the top of the panel). 
}
\end{figure*}

To study the possibility of noreciprocal MSWs propagation in a magnetic multilayered structure with asymmetry in the geometrical characteristics of the magnetic constituents, we have chosen a bilayer structure consisting of two dipolarly coupled layers of the ferromagnetic alloy galfenol separated by a copper layer, specifically Fe$_{81}$Ga$_{19}$(7\,nm)/Cu(5\,nm)/Fe$_{81}$Ga$_{19}$(4\,nm), grown by sputter deposition on a (100)-GaAs substrate.
In the following, the 7-nm galfenol layer is referred to as the top layer, and the 4-nm layer adjacent to the substrate -- as the bottom layer (inset in Fig.~\ref{fig:3Lay}\,a).
The top layer is protected by Al~(3\,nm) and SiO$_2$(120\,nm) capping layers.
The preparation and characterization of the structure are described in details elsewhere~\cite{DanilovPRB:2018}.
Furthermore, a single 20\,nm film of galfenol on a (100)-GaAs substrate was deposited with the same capped layers as in the case of bilayer structure~\cite{bowe2017magnetisation}.
The single film has a thickness comparable to the total thickness of the layered structure, and thus it is used as a reference to determine the MSW dispersion in the case of a single film. 
We choose galfenol-based structures because they exhibit low Gilbert damping~\cite{parkes2013magnetostrictive}, high values of magnetization precession lifetime~\cite{scherbakov2019optical, Gerevenkov_PRMat_2021} and long MSW propagation length~\cite{khokhlov2019optical}, which makes galfenol a prospective material for magnonic applications.
According to ferromagnetic resonance measurements, the saturation magnetization of the ferromagnetic layers is estimated as $ \mu_0 M_S = 1.7 $\,T at layer thicknesses of 7 and 20\,nm~\cite{khokhlov2019optical,Gerevenkov_PRMat_2021}. 
The saturation magnetization of the 4\,nm layer is slightly reduced due to the interface with the GaAs substrate~\cite{vaz2008magnetism} and is $ \mu_0 M_S = 1.6 $\,T. 
All galfenol layers demonstrate effective cubic magnetocrystalline anisotropy with the parameter $K_C = 2.7\cdot10^4$~J~m$^{-3}$ and an additional uniaxial growth-induced anisotropy in the film's plane~\cite{atulasimha2011review}.  
The parameters of the uniaxial anisotropy are $K_{U\,t} = -0.3\cdot10^4$~J~m$^{-3}$ for the top layer and $K_{U\,b} = -1.1\cdot10^4$~J~m$^{-3}$ for the bottom layer in the bilayer structure.
$K_U = -1\cdot10^4$~J/m$^3$ for the 20-nm single film~\cite{khokhlov2019optical}. 

The dispersion of thermal MSWs was studied by the Brillouin light scattering (BLS) technique in the backscattering geometry~\cite{bessonov2015magnonic, demidov2016excitation, demidov2017chemical} (inset in Fig.~\ref{fig:3Lay}\,a).
A single mode laser with a wavelength of 532\,nm is focused on the sample surface into a spot with a diameter of 50\,$\mu$m using an objective lens. 
Frequency resolution is achieved using a six-pass Fabry-Perot interferometer. 
The resolution in the wave vectors is achieved by tilting the optical axis of the detection beam from the normal to the film's plane in the range of angles $\theta = 0 - 70$\,$^\circ$. 
The corresponding range of wavenumbers is $k_y = 0 - 22\cdot10^4$\,rad~cm$^{-1}$.
All experiments are carried out at room temperature in an external magnetic field of $\mu_0 H_{\mathrm{ext}} = 100$\,mT applied in the sample plane and perpendicularly to the registered wavevector.
The orientation of $\mathbf{H}_{ext}$ is chosen either along the easy (EA) or hard (HA) magnetization axes in the film's plane~(see details in Sec.~II Suppl. Materials~\cite{Supplemental}). 

\section*{Results and discussion}
Figure~\ref{fig:3Lay}\,a presents a typical BLS spectrum of the bilayer structure obtained at $\theta = 35\,^\circ$, $k_y~=~13.5~\cdot~10^4$\,rad~cm$^{-1}$. 
Two peaks are visible in the Stokes and anti-Stokes parts of the spectrum with central frequencies $f$ of 11 and 18\,GHz. 
The experimental dependencies of $f$ on the wavenumber $k_y$ are obtained by approximating the peaks in the BLS spectra using Lorentz functions (shown by symbols in Fig.~\ref{fig:3Lay}\,b~and~c). 
$f$ of the high frequency peak increases with wavenumber, while the lower one demonstrates only a weak dependence of its position on $k_y$.
This behavior is observed for both directions of $\textbf{H}_{\mathrm{ext}}$, parallel to HA and EA.
To reveal the origin of the two peaks in the BLS spectra, we compare the dispersions obtained with those in the 20-nm single galfenol film shown in Fig.~\ref{fig:3Lay}\,d.
For the single film, the BLS spectra also contain two peaks. 
The frequency of one peak increases with $k_y$, and thus can be readily assigned to the magnetostatic surface spin wave (MSSW). 
The frequency of the second peak is weakly dependent on the wavenumber, and corresponds to the spin wave resonance (SWR). 
In the studied range of $k_y$, the frequencies of the SWR are higher than those of the MSSW, according to the typical properties of these waves in single magnetic layers~\cite{song2021nonreciprocal}. 
In striking contrast, in the bilayer structure, the high frequency peak shows the stronger dependence on $k_y$. 
Furthermore, at $k_y=0$, the frequencies of both dispersion branches coincide, while they differ in the single layer. 
Therefore, the two dispersion branches in the bilayer structure cannot be explained in terms of MSSWs and SWRs. 
Instead, the observed dispersions indicate the formation of the collective magnetization dynamics of two ferromagnetic layers and are described in terms of the collective acoustic and optical modes (Fig.~\ref{fig:3Lay}\,b and c).
The formation of these modes was demonstrated in~\cite{grunberg1986layered, kueny1984magnons} and a theoretical description is given in~\cite{camley1983magnetic,grunberg1980magnetostatic}. 
It should be noted, that the dispersions of MSW in the multilayered structure do not coincide with the dispersions of its individual ferromagnetic constituents (see Sec.~IV Suppl. Materials~\cite{Supplemental}). 

\begin{figure*}
\centering
\includegraphics[width=0.95\linewidth]{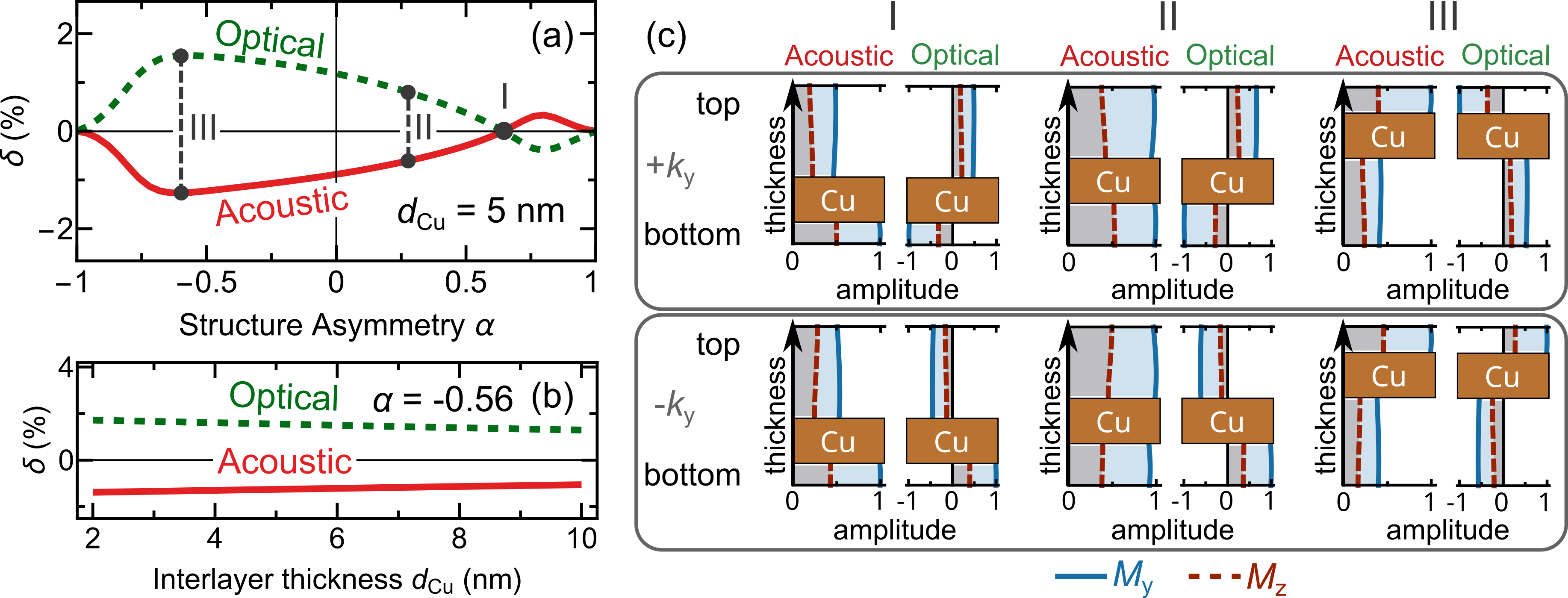}
\caption{\label{fig:NR} 
Nonreciprocity parameter $\delta$ from the model vs. structure asymmetry $\alpha$ (a) and interlayer thickness $d_{\mathrm{Cu}}$ (b) for acoustic (red line) and optical (green line) branches of dispersion.
(c) Distributions of the amplitudes of in-plane $M_y$ and out-of-plane $M_z$ magnetization components over the thickness of the structure for EA the acoustic and optical branches for structures with minimal~(I), maximal~(III) $\delta$ and parameters of bilayer structure under the experiment~(II).
Data are given for the value of $|k_y| = 22 \cdot 10^4$\,rad~cm$^{-1}$.
}
\end{figure*}

To substantiate the attribution of the two peaks in the BLS spectra to scattering by the acoustic and optical modes, we used the theoretical approach developed in~\cite{rado1988theory, hillebrands1990spin, carlotti1999brillouin}.
The details of the theoretical calculations are presented in Sec.~I-III Suppl. Materials~\cite{Supplemental}.
The fit for the single film using the model is shown by lines in Fig.~\ref{fig:3Lay}\,d.
Because of the equality of frequencies for waves propagating in the opposite directions, i.e., zero nonreciprocity in frequencies (see Sec.~V Suppl. Materials~\cite{Supplemental}) for the single galfenol layer, we use the same boundary conditions for all interfaces in the subsequent analysis.
The boundary conditions in the model introduced in the form of in-plane $\zeta$ and out-of-plane $\xi$ pinning parameters. 
For the single film sample, the best fit is achieved at values of the parameters $\xi = -46.6$ and 48.6\,nm, $\zeta = 6.2$ and 11.8\,nm for $\mathbf{H}_{ext}$ along EA and HA, respectively.
The parameter of exchange stiffness of galfenol from the fit is $A = 8\cdot10^{-12}$\,J~m$^{-1}$, which is comparable to the value for pure iron~\cite{niitsu2020temperature}. 

The obtained exchange stiffness parameter makes it possible to approximate the dispersion dependencies of the bilayer structure (lines in Fig.~\ref{fig:3Lay}\,b~and~c). 
Good agreement with the experimental results supporting assignment of the peaks in the BLS spectra to the coupled modes is achieved using a model that takes into account only the dipolar coupling between the layers~\cite{carlotti1999brillouin}. 
Following the single layer case, for the bilayer structure, we assume equal pinning parameters for the external interfaces of the whole bilayer structure. 
The calculation results (see Sec.~VI Suppl. Materials~\cite{Supplemental}) show that a change in the pinning parameters at the interfaces with the Cu interlayer leads only to a quantitative change in the results. 
Since a change in the pinning at the interfaces with copper does not change the simulation results qualitatively, we assume the same pinning parameters for all interfaces further for simplicity.
The values of the pinning parameters are $\xi = -1.5 \cdot 10^5$ and $7.5 \cdot 10^3$\,nm, $\zeta = 88$ and $3.6 \cdot 10^3$\,nm for $\mathbf{H}_{ext}$ along EA and HA, respectively.
Changes in pinning parameters $\xi$ and $\zeta$ compared to the single-film case can be associated with significantly thinner layers in the multilayered structure and the greater influence of interfaces on spin dynamics~\cite{wang2019spin}.
 
Having identified the origin of the peaks in the BLS spectra of the bilayer structure, we examine do the coupled modes possess nonreciprocity. 
Indeed, comparing the frequencies at $\pm k_y$, one notices differences between them.
To quantify nonreciprocity, we introduce the dimensionless parameter $\delta(k_y) = 2[f(k_y)-f(-k_y)]/[f(k_y)+f(-k_y)]$.
In the single galfenol film $\delta$ is equal to zero over the range of studied $k_y$ both in experiments and in the model, confirming that the interfaces alone do not introduce nonreciprocity for such a film~(see Sec.~V Suppl. Materials~\cite{Supplemental} for details of extracting $\delta$ from the experimental data). 
In the bilayer structure, in contrast, nonzero $\delta$ is found for both the acoustic and the optical modes, as shown by dots in Fig.~\ref{fig:3Lay}\,e~and~f for magnetization along EA and HA.
The nonreciprocity $\delta(k_y)$ increasing with $k_y$ and possessing the opposite slope for the acoustic and optical branches is clearly seen when the two spectral peaks in the BLS signal are fully resolved, i.e. when $k_y$ exceeds $\approx10^4$~rad~cm$^{-1}$ (see non-shadowed areas in Fig.~\ref{fig:3Lay}\,e~and~f).
Importantly, the model also predicts the presence of nonreciprocity in the bilayer structure.
The results of the model are shown by lines and are in qualitative agreement with the experiment. 

Such a dependence of nonreciprocity on the wavenumber has been previously found for the case of identical dipole-coupled Co layers with antiparallel magnetizations~\cite{franco2020enhancement}. 
In our experiments, however, the magnetizations of the top and bottom layers are parallel to each other, and thus a different source of nonreciprocity should be identified.
Interface-induced nonreciprocity can be excluded as well, as the single film does not demonstrate nonreciprocity in experiments, and identical boundary conditions for the bilayer structure were used in the model, as discussed above.
No exchange coupling between the layers, which is another possible source of nonreciprocity, is present. 
Thus, we argue that the nonreciprocity observed in the experiments and reproduced in the model is caused by the geometric asymmetry of the structure, i.e. by the difference in the thicknesses of ferromagetic layers, and by the difference in magnetic parameters of the layer. 

\subsection*{Theoretical analysis}

\begin{figure*}
\centering
\includegraphics[width=0.8 \linewidth]{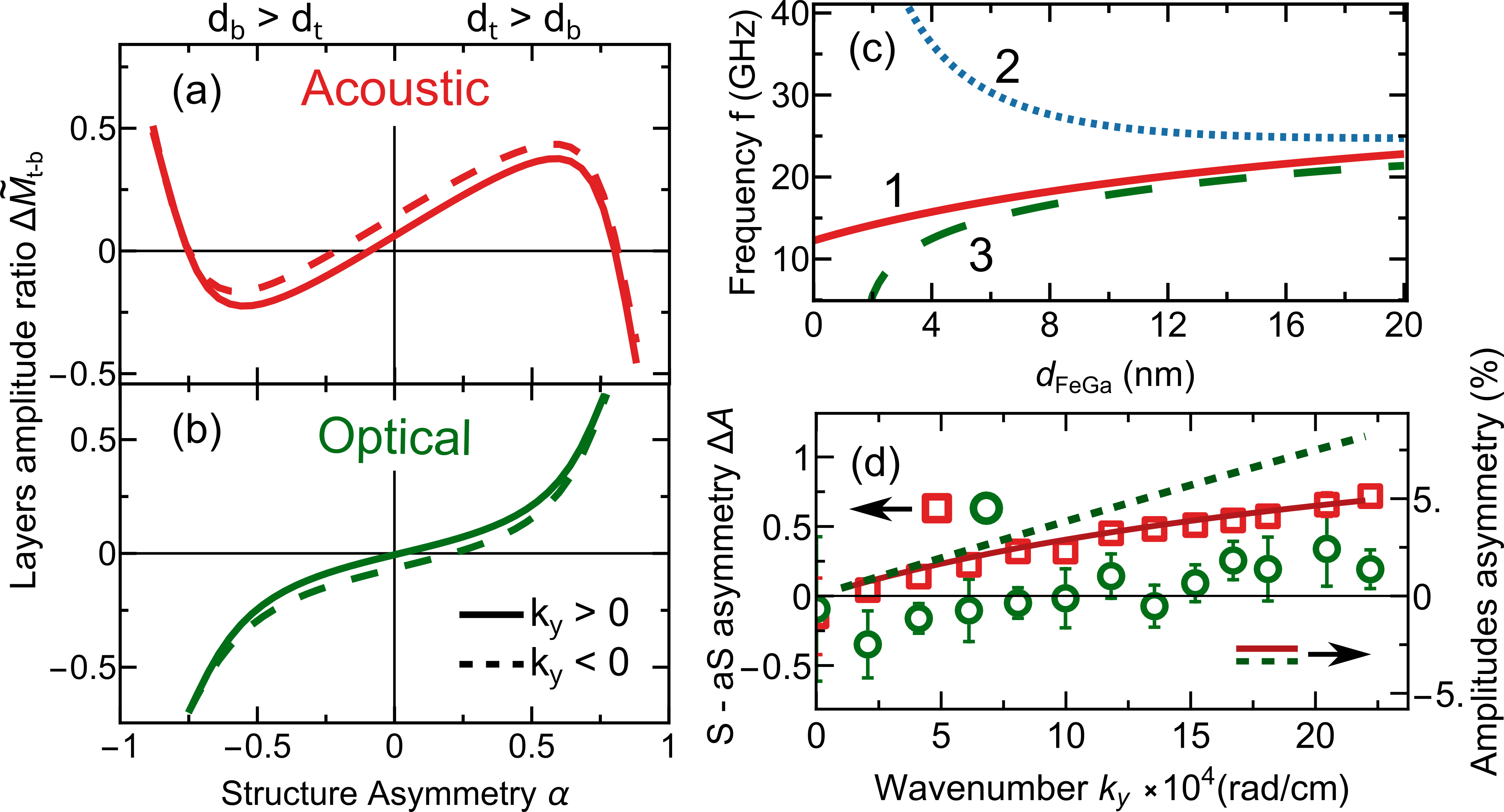}
\caption{\label{fig:El}
(a and b) Relationship in the amplitudes of the magnetization dynamics integrated over the layer thicknesses between two ferromagnetic layers as a function of the geometric asymmetry for the (a) acoustic and (b) optical modes.
Data for a wave propagating in the positive and negative directions are shown by solid and dashed lines, respectively.
(c) Frequency of the magnetostatic surface mode vs. thickness of the single galfenol layer at $k_y = 22 \cdot 10^4$\,rad~cm$^{-1}$.
Line 1 shows frequency behavior without considering exchange interaction and interface anisotropy obtained from the theory~\cite{mahmoud2020introduction}.
Lines 2 and 3 show frequency behavior obtained from model using in this work at pining parameters $\xi = \zeta = -100$\,nm and $\xi = \zeta = 10$\,nm, respectively.
(d) Asymmetry between the BLS peak amplitudes (markers) and between the $M_z$ values on the top of the structure at $k_y = \pm 22 \cdot 10^4$\,rad~cm$^{-1}$ (lines).
Data for the acoustic branch are shown with square markers and solid line, and data for the optical branch are shown with circles and dotted line.
}
\end{figure*} 

To determine the impact of geometric asymmetry, we independently vary in the model the thickness of the nonmagnetic spacer $d_{\mathrm{Cu}}$ and the parameter of the geometrical asymmetry of the structure $\alpha=(d_{t}-d_{b})/(d_{t}+d_{b})$, where $d_{t}$ and $d_{b}$ are the thicknesses of the top and bottom ferromagnetic layers, respectively.
Figure~\ref{fig:NR}\,a shows the dependence $\delta (\alpha)$ for both collective spin wave modes at $|k_y| = 22 \cdot 10^4$\,rad~cm$^{-1}$, $d_{\mathrm{Cu}} = 5$\,nm, and $\mathbf{H}_{ext}$ directed along EA.
The values of $\delta$ for the acoustic and optical branches of the dispersion are of opposite signs for all $\alpha$.
The modulus of $\delta$ for the optical branch is somewhat higher than for the acoustic one for the entire range of $\alpha$.
At a certain value of $\alpha$ the nonreciprocity inverts its sign for both branches (see the point I in Fig.~\ref{fig:NR}\,a).
We note, that point I is shifted relative to the completely symmetric structure with $\alpha = 0$.
For the structure with equal magnetic parameters of the two ferromagnetic layers, no nonreciprocity is found in a symmetric structure for both dispersion branches (see Sec.~VII in Suppl. Material~\cite{Supplemental}) in agreement with previous reports~\cite{zhang1987spin}.
In this case, the asymmetric structures with opposite sign of $\alpha$ show equal absolute values of $\delta$.
$\delta$ can reach tens of percent, if the structure with appropriate pinning parameters on interfaces and geometrical asymmetry will be prepared~(Fig.~S.5\,a in Sec.~VIII in Suppl. Material~\cite{Supplemental}).
Figure~\ref{fig:NR}\,a shows that the geometric asymmetry gives a substantial contribution to the nonreciprocity, and it is at least comparable to the contribution of the difference in the magnetic parameters.
As the asymmetry is tuned from the point I towards -1 or +1, $\delta(\alpha)$ increases and reaches the extreme points marked as III in Fig.~\ref{fig:NR}\,a.
The values of asymmetry at which maxima of $|\delta|$ are observed, are the same for both branches, and depend on the pinning parameters $\zeta$ and $\xi$. 
The asymmetry of the structure under study is marked by point II in Fig.~\ref{fig:NR}\,a.

Figure~\ref{fig:NR}\,b shows the dependence of $\delta$ on $d_{\mathrm{Cu}}$ for the asymmetry value at which the nonreciprocity is maximal ($\alpha = -0.56$). 
The dependence of the nonreciprocity on the nonmagnetic interlayer thickness has the same character for all values of $\alpha$.
Decreasing $d_{\mathrm{Cu}}$ in the range from 10 to 2\,nm gives only a weak increase of the nonreciprocity confirming the dominating role of structural asymmetry in the emergence of nonreciprocity.
With a further decrease of $d_{\mathrm{Cu}}$, it would be necessary to take into account the exchange interaction between the ferromagnetic layers~\cite{heinrich1993ultrathin}, which is discussed by other authors~\cite{shiota2020tunable, franco2020enhancement} and is beyond the scope of this work.

To illustrate the physical picture underlying the observed and predicted nonreciprocity, we show in Fig.~\ref{fig:NR}\,c the distributions of the amplitudes of the dynamic in-plane $M_y$ and out-of-plane $M_z$ magnetization components over the thickness of the structure for various values of nonreciprocity (points I – III in Fig.~\ref{fig:NR}\,a).
Distributions are given for the optical and acoustic branches for waves propagating in opposite directions.
Based on these distributions, we compute the shift $\Delta\tilde M_{t-b}=(\tilde M_t-\tilde M_b)/(\tilde M_t+\tilde M_b)$ of the amplitudes of the modes $\tilde M_{t,b}=\sqrt{\tilde M^2_{y(t,b)}+\tilde M^2_{z(t,b)}}$ integrated over the thickness of the layer towards the top or the bottom layer.
Figures~\ref{fig:El}\,a and b show the shift as a function of $\alpha$ for the parameters of the structure under study.  
The following features should be noted:
\begin{itemize}
  \item Positive slops of $\Delta\tilde M_{t-b}(\alpha)$ indicate that the modes show localization in a thicker layer, except for the case of large $|\alpha|$ for the acoustic branch.
  \item The vertical relative displacement of $\Delta\tilde M_{t-b}(\alpha)$ for positive and negative $k_y$ (compare the solid and dashed lines in Fig.~\ref{fig:El}\,a~and~b) shows that the maximum of the mode amplitude is additionally shifted to the upper or the lower ferromagnetic layer, depending on the sign of $k_y$.
  \item This shift of the amplitude maxima dependent of the sign of $k_y$ appears to be opposite for the acoustic and optical branches (compare Figs.~\ref{fig:El}\,a and b).
\end{itemize}
The combination of the first two features suggest that the nonreciprocity induced by the geometrical asymmetry $\alpha$ originates from a stronger or weaker localization in the thicker layer, depending on the $k_y$ sign.
The third feature explains the opposite sign of nonreciprocity observed for the acoustic and optical branches.

Fig.~\ref{fig:El}\,c shows the frequency change with thickness for a single galfenol layer.
Line 1 shows the frequency of the pure surface magnetostatic mode without considering exchange interaction and interface anisotropy obtained from the theory~\cite{mahmoud2020introduction}.
One sees that, even in this simple case, there is a nonzero change in frequency with the thickness of the thin films.
Thus, the different degree of localization of the coupled spin waves mode in thicker and thinner layer controlled by the sign of $k_y$ (Fig.~\ref{fig:El}\,a~and~b) affects the frequency of this mode.
This effect is enhanced if one accounts for the interfacial pinning and exchange interaction within the ferromagnetic layer.
Lines 2 and 3 in Fig.~\ref{fig:El}\,c demonstrate thickness dependences of the MSW frequency for different pining parameters.
As one can see, depending on the pinning parameters, the variation of the frequency with thickness becomes more potent pronounced and can exhibit different slopes.
Such a behavior has been experimentally demonstrated for iron films on various substrates~\cite{hillebrands1987situ,carlotti1999brillouin}.
Importantly, for layers thicker than 20\,nm, the frequency is almost independent of the thickness (Fig.~\ref{fig:El}\,c).
In the model, an increase in the total thickness of two ferromagnetic layers leads to a decrease in nonreciprocity (Sec.~IX in Suppl. Material~\cite{Supplemental}).
As the layer thickness decreases below $\approx$5~nm, the change of the frequency with the thickness becomes drastic, which can prevent the formation of coupled modes in structures with a high value of $\alpha$, as observed in our calculations for $|\alpha|>0.8$, indeed.

Finally, we address the Stokes / anti-Stokes amplitude asymmetry introduced as $\Delta A = (A_{S}-A_{aS})/(A_{S}+A_{aS})$, where $A_{S}$ and $A_{aS}$ are the peak amplitudes in the Stokes and anti-Stokes parts of the BLS spectrum, obtained from the fit.
The Stokes / anti-Stokes amplitude asymmetry $\Delta A$ for the acoustic and optical branches both increase with increasing $|k_y|$ as shown in Fig.~\ref{fig:El}\,d.
To explain this fact, we analyzed amplitudes and related Stokes / anti-Stokes asymmetry $\Delta A$ for both branches at the top interface of the structure predicted by the model, as the magnetization dynamics at this interface gives the dominating contribution to the BLS signal, especially for the large $k_y$.
The results for the acoustic and optical modes are shown in Fig.~\ref{fig:El}\,d by solid and dashed lines, respectively.
The data from the model demonstrate qualitative agreement with the experiment.
This behavior can be explained by the distribution of the amplitudes within each layer.
Within one layer, both modes have an amplitude distribution typical for a surface MSW, that is, the amplitude's maximum shifts to the bottom interface of the layer for positive values of $k_y$ and to the top interface for negative ones.
As a result, the Stokes / anti-Stokes asymmetry $\Delta A$ in the BLS experiment appears to be of the same sign for both modes.
Thus, while the nonreciprocity $\delta$ stems from the predominant localization of the modes in one of the layers of the bilayer structure, the Stokes / anti-Stokes asymmetry is governed by the shift of the spin wave amplitudes within its top layer. 

\section*{Conclusions}
In conclusion, we have experimentally demonstrated the nonreciprocal behavior of the collective spin wave modes in the system of two parallel magnetized ferromagnetic layers of galfenol separated by a nonmagnetic copper spacer with pure dipolar coupling between them.
The analysis shows that the geometric asymmetry of the structure, i.e. difference in thicknesses of the ferromagnetic layers, gives a pronounced contribution to the observed nonreciprocity, which appears to be comparable to the contribution because of the differences in the magnetic parameters of the magnetic layers.
Modeling shows that the nonreciprocity in the geometrically asymmetric structure stems from shift of the spin wave mode maxima towards a thicker or thinner layer controlled by the sign and value of $k_y$, and from the dependence of the spin waves frequencies on the thin layers thickness.
Total nonreciprocity achieved in the experiments reaches 2\,\% for $k=22 \cdot 10^4$\,rad~cm$^{-1}$.
However, it can be increased up to tens of percent in the structure with appropriate pinning parameters on interfaces and geometrical asymmetry.

Although this nonreciprocity is somewhat lower than the one achieved because of the difference in saturation magnetization between the layers of multilayered structures~\cite{grassi2020slow, song2020backward}, its value is sufficient to have an impact on the total spin waves' nonreciprocity in multilayered structures. 
Our analysis further suggests that there is an optimum ratio between the layer thicknesses supporting both the formation of the coupled modes and their nonreciprocity, which is also controlled by the spins pinning at interfaces.
Thus, tailored dispersion and nonreciprocity of the coupled modes can be obtained by fabricating a structure with certain values of the pining parameters~\cite{dieny2017perpendicular, wastlbauer2005structural} and geometrical asymmetry.
Additionally, as shown in recent work~\cite{shelukhin2022spin}, the pinning parameters and thus the behavior of spin waves can be controlled at ultrashort time scales.
Furthermore, the considered bilayer structure supports optical excitation of coherent coupled modes \cite{danilov2018spinpumping}, in which case mutual spin pumping may contribute to nonreciprocity.
Therefore, the demonstrated geometrical control of spin wave nonreciprocity paves the way to the design of magnonic logic gates based on predicted and controllable dispersion of MSWs~\cite{mahmoud2020introduction}.

\begin{acknowledgments}
The authors thank S. O. Demokritov for insightful discussions. 
The work of P.I.G., A.M.K. and N.E.Kh. is supported by RSF (project 20-12-00309, \href{https://rscf.ru/en/project/20-12-00309/}{https://rscf.ru/en/project/20-12-00309/}).
V.D.B. acknowledges RFBR (Grant 19-32-50141).
V. D. B., V. S. T. and A. V. T. acknowledge the Russian Government Program “Spin” no. 122021000036-3.
\end{acknowledgments}

\normalem
\bibliography{ref}

\end{document}